# On the growth and form of spherulites

László Gránásy[1]*, Tamás Pusztai[1], György Tegze[1], James A. Warren[2], Jack F. Douglas[3]

[1]*Research Institute for Solid State Physics and Optics, P.O. Box 49, H-1525 Budapest, Hungary;*
[2]*Metallurgy and* [3]*Polymers Divisions, National Institute of Standards and Technology, Gaithersburg, Maryland 20899*

Many structural materials (metal alloys, polymers, minerals, etc.) are formed by quenching liquids into crystalline solids. This highly non-equilibrium process often leads to polycrystalline growth patterns that are broadly termed 'spherulites' because of their large-scale average spherical shape. Despite the prevalence and practical importance of spherulite formation, only rather qualitative concepts of this phenomenon exist. The present work explains the growth and form of these fundamental condensed matter structures on the basis of a unified field theoretic approach. Our phase field model is the first to incorporate the essential ingredients for this type crystal growth: anisotropies in both the surface energy and interface mobilities that are responsible for needle-like growth, trapping of local orientational order due to either static heterogeneities (impurities) or dynamic heterogeneities in highly supercooled liquids, and a preferred relative grain orientation induced by a misorientation-dependent grain boundary energy. Our calculations indicate that the diversity of spherulite growth forms arises from a competition between the ordering effect of discrete local crystallographic symmetries and the randomization of the local crystallographic orientation that accompanies crystal grain nucleation at the growth front (growth front nucleation or GFN). The large-scale isotropy of spherulitic growth arises from the predominance of GFN.

## I. INTRODUCTION

Spherulites are ubiquitous in solids formed under highly non-equilibrium conditions [1]. They are observed in a wide range of metallurgical alloys, in pure Se [2,3], in oxide and metallic glasses [4,5], mineral aggregates and volcanic rocks [6,7], polymers [1,8], liquid crystals [9], simple organic liquids [10], and diverse biological molecules [11]. Many everyday materials, ranging from plastic grocery bags to airplane wings and cast iron supporting beams for highway bridges, are fabricated by freezing liquids into polycrystalline solids containing these structures. The properties and failure characteristics of these materials depend strongly on their microstructure, but the factors that determine this microstructure remain poorly understood [1].

While the term 'spherulite' suggests a nearly spherical shape (circular shape in two dimensions where the term spherulite is still employed), this term is used in a broader sense of densely branched, polycrystalline solidification patterns [2,9,12–21]. Spherulitic patterns exhibit a diversity of forms and representative patterns are shown in Fig. 1.

Experimental studies performed over the last century indicate that there are two main categories of spherulites [20,21]. Category 1 spherulites grow radially from the nucleation site, branching intermittently to maintain a space filling character (Fig. 2). In contrast, category 2 spherulites grow initially as thread-like fibers, subsequently forming new grains at the growth front (Fig. 2). This branching of the crystallization pattern ultimately leads to a crystal 'sheaf' that increasingly splays out during growth. At still longer times, these sheaves develop two 'eyes' (uncrystallized regions) on each side of the primary nucleation site [see Fig. 1(h)]. Ultimately, this type of spherulite settles down into a spherical growth pattern, with eye structures apparent in its core region. In some materials, both categories of spherulite occur in the same material under the same nominal thermodynamic conditions [Fig. 1(i)].

While the widely different systems indicated in Fig. 1 surely involve disparate molecular scale dynamical processes, the evident similarities of their morphologies (tendency for space filling, polycrystallinity, elongated fiber-like grains, etc.) suggest that a general coarse-grained description of this type of pattern formation can be formulated.

While there is no generally accepted theory of spherulite crystallization, a number of phenomenological models and necessary physical conditions for this process have been suggested [1,8–10,22]. The most prevalent conception of their origin is the qualitative model of Keith and Padden [13], in which the presence of static heterogeneities (impurities or molecular defects and mass polydispersity in polymeric materials) lead to a rejection of these components from the growth front to form channels similar to those found in eutectics. The observation of spherulitic growth in highly *pure* liquids by Magill and others [1–3], however, indicates that this cannot be a general explanation of this growth form. Magill, and others preceding him [23], have emphasized that a critically large viscosity, characteristic of high supercooling, seems to be required for spherulites to form. The occurrence of "secondary" nucleation at the growth front (similar to "sympathetic" nucleation observed during solid state precipitation [24] or "double nucleation" in the biological literature [25–28]) has also been emphasized as an essential feature of spherulite formation in polymeric fluids [29]. Random lamellar branching with preferred crystallographic misfit is also expected to play an important role [1,15]. Recent experimental studies of spherulitic growth in thin polymer films by atomic force microscopy strongly support these views [30].

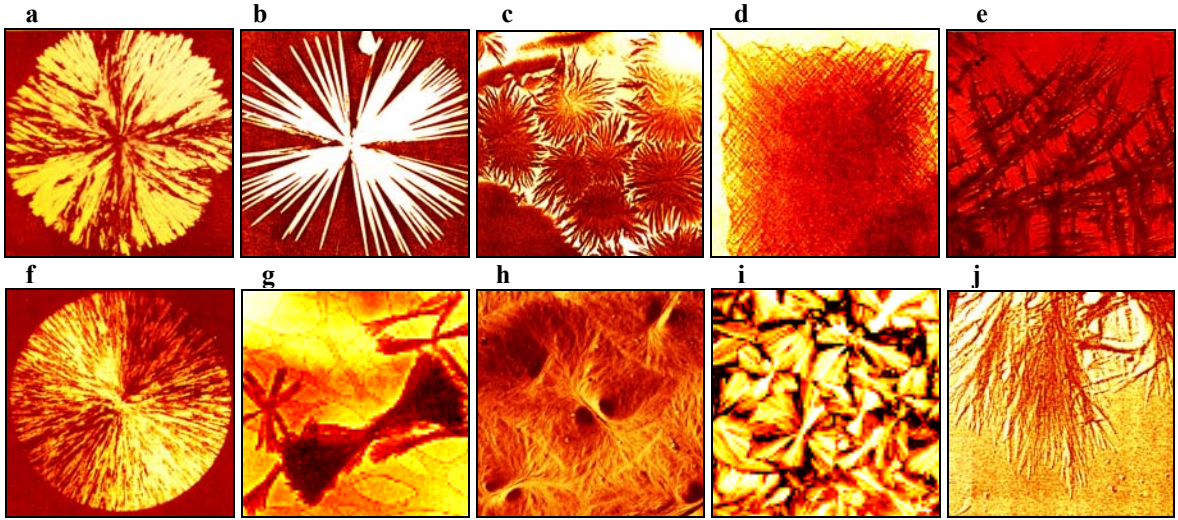

FIG. 1. Various spherulitic morphologies. (a) Densely branched spherulite formed in a blend of isotactic and atactic polypropylene (Ref. 12). (b) 'Spiky spherulite' grown in malonamide-d-tartatic acid mixture (Ref. 13). (c) Arboresque spherulites forming in polypropylene film (Ref. 14), (d) and (e) 'Quadrites' formed by nearly rectangular branching in isotactic polypropylene (Refs. 15, 16). (f) Spherulite formed in pure Se (Ref. 2). (g) Crystal sheaves in pyromellitic dianhydride-oxydianilin poly(imid) layer (Ref. 17). (h)Typical category 2 spherulites (a thin film of polybutene) with two "eyes" on the sides of the nucleus (Ref. 18). (i) Multi-sheave / early spherulite structure formed in dilute long n-alkane blend (Ref. 19). (j) Arboresque growth form in polyglycine (Ref. 20). To improve the contrast/visibility of the experimental pictures, false colors were applied. The linear size of the panels are (a) 220 μm, (b) 960 μm, (c) 2.4 mm, (d) 2.5 μm, (e) 7.6 μm, (f) 550 μm, (g) 2.5 μm, (h) 20 μm, (i) 250 μm, and (j) 1.7 μm, respectively.

Our previous work [31] proposed a unified description of the origin of polycrystalline growth. We established that polycrystalline growth generally arises from the quenching of orientational defects that can arise from *either* static heterogeneities (impurities) or dynamic heterogeneities intrinsic to supercooled liquids. We termed this secondary nucleation of crystal grains at the crystal growth front as *growth front nucleation* (GFN). Both types of disorder yield strikingly similar effects on crystallization morphologies [31]. Thus, we can expect spherulite formation to occur in both highly impure and pure supercooled fluids. Spatial heterogeneities due to phase separation can provide another source of static disorder giving rise to spherulitic growth [32,33]. Given this duality between dynamic and static heterogeneities, we focus herein on polycrystalline growth in particulate-free supercooled liquids. In this case, the glass-forming nature of the fluid is found to play a key role in the spherulite formation process and we briefly review some of the essential aspects of this phenomenon.

It is now appreciated that highly supercooled liquids are characterized by the presence of long-lived dynamic heterogeneities. These heterogeneities are associated with the formation of regions within the fluid that have either a much higher or much lower mobility relative to a simple fluid in which particles exhibit Brownian motion [1,34–36]. These nanoscale heterogeneities persist on timescales of the order of the stress relaxation time, which can be minutes near the glass transition and eons at lower temperatures. The presence of such transient heterogeneities leads to dramatic effects on the dynamics of supercooled liquids [37–41].

Dynamic heterogeneity has numerous consequences for the transport properties of these complex fluids. The most important transport properties of relevance to crystallization are the shear viscosity ($\eta$) and the molecular mobilities determined by the translational ($D_{tr}$) and rotational diffusion ($D_{rot}$) coefficients. These diffusion coefficients characterize the rate of molecular translation and rotation, directly controlling the manner that molecules attach and align with the growing crystal [34,39–41]. It is a common property of highly supercooled liquids that the ratio of the rotational and translational diffusion coefficients ($\chi = D_{rot} / D_{tr}$) decreases sharply (by orders of magnitude) from their nearly constant high temperature values ($\chi_0$) [34,37–42]. This 'decoupling' phenomenon means that molecules translate increasingly large distances before they rotationally decorrelate from their initial orientation [34,37–41].

Recently we demonstrated that a drop in $\chi$, characteristic of highly supercooled liquids, enhances the growth of new grains as misoriented crystal regions at the liquid-solid interface have difficulty aligning with the parent crystal. In other words, polycrystalline growth will arise if the reorientation of molecules is slow relative to the interface propagation. This argument implies that static heterogeneities and the mobility asymmetry ($\chi \ll \chi_0$) of supercooled liquids should give rise to a common tendency towards polycrystalline growth. It is emphasized that our simulations do not model the nanoscale dynamic heterogeneities explicitly, but instead model the consequences of these heterogeneities on molecular transport, as is appropriate for a coarse-grained model.

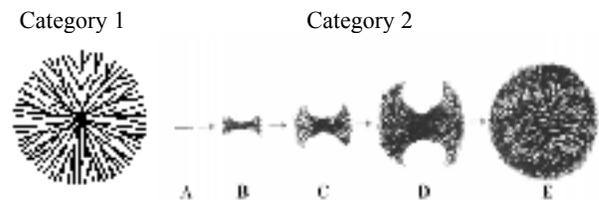

FIG. 2. Concepts for the formation of category 1 and 2 spherulites. From left to right: Category 1 spherulite formed via central multidirectional growth. Formation of category 2 spherulite from a folded-chain single crystal (A) to the fully developed spherulite (E) via unidirectional growth and low angle branching (Ref. 15). Note that the latter mechanism may lead to the formation of two 'eyes' (uncrystallized holes) on the sides of the nucleation site.

Here we present a unified model of polycrystalline solidification that incorporates the essential ingredients needed to describe complex growth morphologies and explore its ability to describe polycrystalline spherulites.

## II. PHASE FIELD THEORY WITH CRYSTALLOGRAPHIC BRANCHING

Our two-dimensional phase field theory builds on the phase field models of primary nucleation of crystals from the melt [43] and multigrain solidification [43,44], which incorporate the diffusional instabilities and crystal anisotropies of the interface free energy and molecule-attachment kinetics, and the possibility for trapping orientational defects into the solid. This model has been successfully applied to describe transformation kinetics in alloys [43] and the interaction of particulate additives with dendrites [45]. (Recent reviews on the phase field technique and its application to polycrystalline solidification are available in Refs. 46–48.)

The novel aspect of the approach used in the present paper is the introduction of *branching with a fixed crystallographic misorientation*, realized through an orientation-dependent grain boundary energy. The combination of these essential factors provides a general model of polycrystalline solidification, suitable to describe the formation of complex polycrystalline patterns, in particular the growth and form of spherulites.

The local state of matter is characterized by the phase field $\phi$. This order parameter describes the extent of structural change during freezing and melting. The other basic field variables are the chemical composition $c$ and the normalized orientation field $\theta$ [43], where $\theta$ specifies the orientation of crystal planes in the laboratory frame. The free energy $F$ consists of various contributions that will be discussed below:

$$F = \int d^3r \left\{ \frac{\alpha_0^2 T}{2} s^2(\vartheta,\theta)|\nabla\phi|^2 + f(\phi,c,T) + f_{ori}(|\nabla\theta|) \right\} \quad (2.1)$$

where

$$f(\phi,c,T) = w(c)Tg(\phi) + [1-p(\phi)]f_S(c,T) + p(\phi)f_L(c,T)$$

$$\alpha_0^2 = \frac{6\sqrt{2}\gamma_{A,B}\delta_{A,B}}{T_{A,B}}, \quad w(c) = (1-c)w_A + cw_B, \quad w_{A,B} = \frac{12\gamma_{A,B}}{\sqrt{2}\delta_{A,B}T_{A,B}},$$

$$g(\phi) = \tfrac{1}{4}\phi^2(1-\phi)^2, \quad g'(\phi) = \phi^3 - \tfrac{3}{2}\phi^2 + \tfrac{1}{2}\phi$$

$$p(\phi) = \phi^3(10-15\phi+6\phi^2), \quad p'(\phi) = 30\phi^2(1-\phi)^2$$

and

$$f_{ori} = [1-p(\phi)]\frac{HT}{2\xi_0}\{xF_0 + (1-x)F_1\}$$

$$F_0 = \begin{cases} |\sin(2\pi m\xi_0|\nabla\theta|)| & \text{for} \quad \xi_0|\nabla\theta| < \dfrac{3}{4m} \\ 1 & \text{otherwise} \end{cases}$$

$$F_1 = \begin{cases} |\sin(2\pi n\xi_0|\nabla\theta|)| & \text{for} \quad \xi_0|\nabla\theta| < \dfrac{1}{4n} \\ 1 & \text{otherwise} \end{cases}$$

$$s(\vartheta,\theta) = 1 + s_0\cos[k(\vartheta - 2\pi\theta/k)], \quad \vartheta = \arctan[(\nabla\phi)_y/(\nabla\phi)_x]$$

$$\theta \in [0,1], \qquad \Xi = \frac{2\pi}{k}\theta$$

Here $\alpha_0$ is a constant, $T$ the temperature, and $\Xi$ is the orientation angle in the laboratory frame. The gradient term for the phase field leads to a diffuse crystal-liquid interface, a feature observed both in experiment [49] and computer simulations [50]. The free energy density $f(\phi,c,T)$ has two minima ($\phi = 0$, and $\phi = 1$, corresponding to the crystalline and liquid phases), whose relative depth is the driving force for crystallization and is a function of both temperature and composition as specified by the free energy densities in the bulk solid and liquid, $f_{S,L}(c,T)$, respectively.

The dependence of the surface energy on orientation of the liquid solid interface is introduced through the function $s(\vartheta,\theta)$, which multiplies the penalty for gradients in $\phi$. As $s$ introduces a misorientation dependence to the surface energy [51], it is possible to introduce favored misorientations through this coefficient. However, it is also possible to introduce misorientation dependencies via a coupling to gradients in $\theta$. Specifically, preferred crystallographic misfits are introduced into our model through the orientational contribution to the free energy density $f_{ori}$, that represents the excess free energy density due to inhomogeneities in crystal orientation in space, in particular the misorientation due to a grain boundary. Its form ensures that $\theta$ takes an essentially constant value (scaled between 0 and 1) in the solid, while in the liquid it fluctuates. The latter feature reflects the local order in the liquid. Orientational ordering takes place at the diffuse interface simultaneously with the structural transition. The free energy of low-angle grain boundaries scales with $HT$, $\xi_0$ is the correlation length of the orientation field, while $p(\phi)$, following the procedure commonly used in phase field theory [52,53], varies smoothly from 0 to 1 as $\phi$ changes from the solid to the liquid. The orientational free energy has two local minima as a function of the angle $\xi_0|\nabla\theta|$, corresponding to no misorientation and a preferred misorientation (Fig. 3). This means that regions with a large enough orientation difference from a neighboring parent crystal will relax towards a finite misorientation. This selection of grain orientation only occurs provided that noise does not disrupt the process. The branching angle and the depth of this metastable minimum of $f_{ori}$ are specified by $m$, $n$ and $x$.

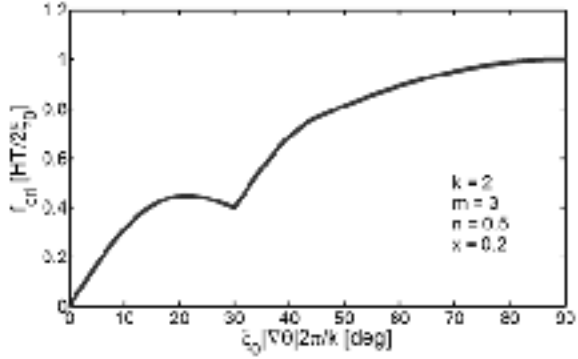

FIG. 3. Orientational free energy $f_{ori}$ as a function of misorientation angle (in degree) for two-fold symmetry ($k = 2$), while $n = \frac{1}{2}$, $m = 3$, and $x = 0.2$. If the neighboring pixel has a smaller misorientation than ~20° (local maximum), it can reduce the free energy by relaxing to the bulk crystal orientation (0°). If misorientation is larger than this, the closest minimum is 30°. So, neighboring pixels of large misorientation tend to relax to 30°, unless fluctuations prevent this. Note that $\theta$ is an angular variable, so the maximum possible misorientation is $\Delta\theta_{max} = 0.5$.

In any real system there will be many preferred (low energy) orientations, a reflection of the underlying crystallographic symmetries. In our illustrative calculations $n = \frac{1}{2}$ has been set, while $m = 1$, 2, and 3 correspond to branching with 90, 45, and 30 degrees, respectively. We note that, with appropriate choice of the parameters ($x = 0$), GFN with random orientation of the new grains [29,48] can also be recovered.

Since we are modeling quasi-two-dimensional systems, the orientation field is simply a scalar, which is suitable for the description of transformations in thin layers of thickness $L$, where along thickness (direction $z$) the system is considered uniform. The true three-dimensional (3D) free energy functional would depend on a 3D vectorial orientation field.

### A. The governing equations

Time evolution is governed by relaxational dynamics and Langevin noise terms are added to model thermal fluctuations [43,48],

$$
\begin{aligned}
\dot{\phi} &= -M_\phi \frac{\delta F}{\delta \phi} = M_\phi \left\{ \nabla\left(\frac{\partial f}{\partial \nabla\phi}\right) - \frac{\partial f}{\partial \phi} \right\} + \zeta_\phi \\
\dot{c} &= \nabla M_c \nabla \frac{\delta F}{\delta c} = \nabla\left\{ Dc(1-c)\nabla\left[\left(\frac{\partial f}{\partial c}\right) - \nabla\left(\frac{\partial f}{\partial \nabla c}\right) - \zeta_j\right] \right\} \\
\dot{\theta} &= -M_\theta \frac{\delta F}{\delta \theta} = M_\theta \left\{ \nabla\left(\frac{\partial f}{\partial \nabla\theta}\right) - \frac{\partial f}{\partial \theta} \right\} + \zeta_\theta
\end{aligned}
\tag{2.2}
$$

where $\zeta_i$ are the appropriate Langevin-noise terms.

The time scales for the three fields are determined by the appropriate mobilities appearing in the equations of motion, and $M_\phi$, $M_c$ and $M_\theta$ are the mobilities associated with coarse-grained equation of motion, which in turn are related to their microscopic counterparts. The mobility $M_c$ is directly proportional to the classic interdiffusion coefficient for a binary mixture. The mobility $M_\phi$ dictates the rate of crystallization, while $M_\theta$ controls the rate at which regions reorient.

As discussed in the introduction, dynamic heterogeneities exist at the nanometer scale, but we do not model these fluctuations directly, as our model is coarse-grained. Since $\chi/\chi_0$ is characteristically small in supercooled liquids, we postulate a corresponding reduction in the ratio of $M_\theta/M_\phi$ to model the average effect of dynamic heterogeneity on global relaxation. This assumption is plausible because these coarse-grained mobilities are functions of their molecular counterparts. Moreover, recent experiment has shown that the rate of crystallization in highly supercooled liquids is proportional to $D_{tr}$, even under decoupling conditions [34,40,41]. In our model, the growth velocity scales linearly with $M_\phi$, so consistency requires $M_\phi \propto D_{tr}$. Since we also expect that $M_\theta \propto D_{rot} \propto 1/\eta$, we arrive at $\chi \propto M_\theta/M_\phi$.

#### 1. Phase field

Using the length and time scales $\xi$ and $\xi^2/D_l$, respectively, where $D_l$ is the chemical diffusion coefficient in the liquid, the dimensionless phase field mobility $m_\phi = m_{\phi 0}\{1 + \delta_l \cos[k(\psi - \theta)]\}$, and $m_{\phi 0} = M_\phi \alpha_l^2 T/D_l$, the following *dimensionless form* emerges

$$
\tilde{\dot{\phi}} = m_\phi \left[
\begin{aligned}
&\tilde{\nabla}(s^2\tilde{\nabla}\phi) - \frac{\partial}{\partial\tilde{x}}\left\{s\frac{\partial s}{\partial\vartheta}\frac{\partial\phi}{\partial\tilde{y}}\right\} + \frac{\partial}{\partial\tilde{y}}\left\{s\frac{\partial s}{\partial\vartheta}\frac{\partial\phi}{\partial\tilde{x}}\right\} \\
&- \xi^2 \frac{w(c)Tg'(\phi) + p'(\phi)\{f_L(c,T) - f_S(c,T) - f_{ori}\}}{\varepsilon_\phi^2 T}
\end{aligned}
\right]
\tag{2.3}
$$

Henceforth quantities with tilde are dimensionless, while prime denotes differentiation with respect to the argument.

#### 2. Concentration field

Following previous works [53,54], we choose the mobility of the concentration field as $M_c = (v_m/RT) D c (1 - c)$, where $v_m$ is the average molar volume, and $D = D_s + (D_l - D_s) p(\phi)$ is the diffusion coefficient. This choice ensures diffusive equation of motion. Since $HT$ is assumed independent of concentration, no coupling to the orientation field emerges. Introducing the reduced diffusion coefficient $\lambda = D/D_l$, the *dimensionless equation of motion* for the concentration field reads as

$$
\tilde{\dot{c}} = \tilde{\nabla}\left\{ \frac{v_m}{RT}\lambda c(1-c)\tilde{\nabla}\left[
\begin{aligned}
&(w_B - w_A)Tg(\phi) + [1 - p(\phi)]\frac{\partial f_S}{\partial c}(c,T) \\
&+ p(\phi)\frac{\partial f_L}{\partial c}(c,T)
\end{aligned}
\right] \right\}
\tag{2.4}
$$

TABLE I. Physical properties of Cu and Ni

|  |  | Cu | Ni |
|---|---|---|---|
| $T_f$ | (K) | 1358 | 1728 |
| $L$ | (J/cm$^3$) | 1728 | 2350 |
| $\gamma$ | (mJ/m$^2$) | 247 | 315 |
| $\delta$ | (nm) | $\sim 1$ | $\sim 1$ |
| $D_l$ | (cm$^2$/s) | $10^{-5}$ | $10^{-5}$ |

### 3. Orientation field

Introducing the dimensionless correlation length of the orientation field $\tilde{\xi}_0 = \xi_0 / \xi$, and defining the dimensionless orientational mobility as $m_\theta = M_\theta \, \xi \, HT / D_l$, the *dimensionless equation of motion* is as follows

$$\dot{\tilde{\theta}} = m_\theta \left[ \begin{array}{l} \tilde{\nabla} \left\{ [1 - p(\phi)] \pi \left[ x \tilde{F}_0 m + (1-x) \tilde{F}_1 n \right] \dfrac{\tilde{\nabla} \theta}{|\tilde{\nabla} \theta|} \right\} \\ - \dfrac{\alpha_0^2}{H\xi} s \dfrac{\partial s}{\partial \theta} \left| \tilde{\nabla} \phi \right|^2 \end{array} \right] \quad (2.5)$$

where

$$\tilde{F}_0 = \begin{cases} \text{sign} \left[ \sin \left( 2\pi m \, \tilde{\xi}_0 \left| \tilde{\nabla} \theta \right| \right) \right] \cos \left( 2\pi m \, \tilde{\xi}_0 \left| \tilde{\nabla} \theta \right| \right) & \text{for} \quad \tilde{\xi}_0 \left| \tilde{\nabla} \theta \right| < \dfrac{3}{4m} \\ 0 & \text{otherwise} \end{cases}$$

$$\tilde{F}_1 = \begin{cases} \text{sign} \left[ \sin \left( 2\pi n \, \tilde{\xi}_0 \left| \tilde{\nabla} \theta \right| \right) \right] \cos \left( 2\pi n \, \tilde{\xi}_0 \left| \tilde{\nabla} \theta \right| \right) & \text{for} \quad \tilde{\xi}_0 \left| \tilde{\nabla} \theta \right| < \dfrac{1}{4n} \\ 0 & \text{otherwise} \end{cases}$$

This form of $f_{\text{ori}}$, and the noise added to the equation of motion ensure that the orientation field $\theta$ is random in space and time in the liquid. This makes possible to quench orientational defects into the solid, leading to polycrystalline growth. Independently, branching with fixed relative misorientation may occur, i.e., sharp (step-like) grain boundaries of fixed orientational misfit (of fixed grain boundary energy) appear.

The second term on the RHS of Eq. (2.5) must be handled with care. It is negligible if the physical interface thickness ($\sim 1$ nm) is used. Due to limitations of computer power, we employ a relatively broad interface compared to those found in metallic alloys. This broad interface leads to artifacts that are not present with thinner interfaces. As a practical matter, we adopt one of the following measures: (a) perform the calculations in the presence of only kinetic anisotropy (then this term is zero); (b) we omit this term.

### 4. The noise

Gaussian noises of amplitude $\zeta = \zeta_s + (\zeta_l - \zeta_s) \, p(\phi)$ are added to the non-conserved fields, where $\zeta_l$ and $\zeta_s$ are the amplitudes in the liquid and solid. The noise has been discretized as described in Ref 55. Its amplitude scales with the spatial and time steps, with the temperature and the film thickness as follows:

$$\zeta' = \zeta \, (\Delta x / \Delta x')\cdot(\Delta t' / \Delta t)^{1/2} \cdot (T'/T)^{1/2} \cdot (L/L')^{1/2}, \quad (2.6)$$

where the primed quantities are for the actual simulation, and those without prime belong to a reference state, in which the noise amplitude was $\zeta$.

As pointed out in Ref. 55, the noise amplitude varies with the volume of the simulation cells. In our quasi-2D system, the cell volume is $V = L \, \Delta x^2$, i.e., it depends on the choice of the layer thickness. In other words, the amplitude of the noise might be regarded as an adjustable variable [56]. In the case of the conserved concentration field, random concentration fluxes were added to the equation of motion [55].

### 5. Numerical solution

The governing equations have been solved numerically using an explicit finite difference scheme. Periodic boundary conditions were used. The time and spatial steps were chosen to ensure stability of our solutions. As the computed morphologies are fundamentally determined by thermal fluctuations at the growth front, convergence to a particular morphology, as we refine the grid and timestep, is possible only in a statistical sense (i.e. the rate of GFN, branching frequency, and the solid fraction inside the solidification envelopes). We note that accurate solutions to the orientation equation require approximately 1/50 of the time step required for the stable solution of the other fields.

A parallel code has been developed that relies on the Message Passing Interface (MPI) protocol and was run on a PC cluster built up at the Research Institute for Solid State Physics and Optics, Budapest, exclusively for phase field calculations. This cluster consists of 75 nodes and a server machine. The present paper is based on computations exceeding 40 CPU-years on a 2GHz processor.

We expect the physical systems of interest to have interface thickness of about 1 nm. Using the algorithms we have implemented for our parallel code, we are constrained to use a significantly thicker interface (about 40 times larger). Indeed, these already ambitious calculations would take about 1,000,000 times longer using a physical value of the interface thickness. However, our own examination of the behavior of the model equations, as well as the experience of many others doing phase field research, imply that the structures obtained by these methods are not only qualitatively correct, but also have predictive power. Specifically, the calculations provide insight and understanding into the mechanisms controlling spherulite formation, as well as demonstrating the factors that influence this type of pattern formation. We are confident that algorithmic developments will ultimately enable quantitative computations of the patterns investigated herein.

### C. Materials and simulation parameters

For specificity, we employ the well-studied, ideal solution phase diagram of the Ni-Cu alloy (for relevant properties see Table I.). This choice is not particularly restrictive, as it is formally equivalent to a pure material [53], where thermal diffusion replaces solute diffusion as the dominant transport mechanism. Moreover, the model is no way restricted to metals as our application to polymer materials below demonstrates. Unless stated otherwise, we fix the temperature to be 1574 K, as in previous studies. The orientation dependence of the molecular attachment kinetics is modeled by $M_\phi = M_{\phi,0}\{1 + \delta_0 \cos[k(\vartheta - 2\pi\theta/k)]\}$. The angle $\psi$ is the inclination of the liquid-solid interface in the laboratory frame and $k$ is the symmetry index. The fiber-like crystallites forming in many of the polymeric matter imply a two-fold symmetry ($k = 2$) and a large kinetic anisotropy, which was chosen as $\delta_0 = 0.995$. A similar approach is used to describe the interfacial free energy: $\gamma = \gamma_0 \, s(\vartheta, \theta)$. Crystal growth is sensitive to both kinetic and interfacial free energy anisotropies, where increasing either yields sharper needle crystal morphologies. Our calculations were performed with supersaturations in the range $0.75 \le S = (c_L - c)/(c_L - c_S) \le 1.2$, where $c_L = 0.466219$, $c_S = 0.399112$ and $c$ are the concentrations at the liquidus, solidus, and the initial homogeneous liquid mixture, respectively.

Since the physical thickness of the interface is in the nanometer range and the typical solidification structures are far larger (μm to mm), a full simulation of polycrystalline solidification from nucleation to particle impingement cannot be performed even with the fastest of the present supercomputers. Since we seek here a qualitative understanding, following other authors [53,54], the interface thickness has been increased by a factor of 20.8, the interface free energy has been divided by 6, while the diffusion coefficient has been increased by a factor of 100. This allows us to follow the life of crystallites from birth to impingement on each other. The dimensionless time and spatial steps were $\Delta t = 4.75\times10^{-6}$ and $\Delta x = 6.25\times10^{-3}$, $\xi$ =2.1×10⁻⁴ cm, $\xi_0/\xi = \Delta x$ and $D_l = 10^{-5}$ cm²/s. Unless stated otherwise, dimensionless mobilities of $m_{\phi\theta l} = 1.0$ and $m_{\theta l} = 360$, and $m_{\theta s} = 0$ were applied, while $D_s = 0$ was taken in the solid. White noises of amplitudes 0.0025, 0.00125, and 0.0375 were used for the three fields $\phi$, $c$ and $\theta$, respectively, except in the nucleation runs, where the phase field noise was enhanced to 0.0125 to speed up the process.

## III. RESULTS

### A. Growth of spherulites

First, we explore the fundamental question "how can a crystal grow as a sphere?" Theoretically, one can grow a "ball" with growth kinetics consistent with simple diffusion (i.e. the radius $R$ of the crystal increases as $t^{1/2}$ with time) at low driving forces (supersaturations). For such a shape, the solute rejected from the growing crystal is incorporated into a boundary layer that extends far into the liquid. However, this situation is essentially never observed in real systems, except as a transient.

At larger driving forces, where the system is far from equilibrium, the liquid-solid interface becomes unstable (the Mullins-Sekerka instability [57]), and the crystallization pattern breaks up into a fingered structure commonly termed "seaweed," [Fig. 4(a)]. The lengthscale of the fingers is determined by a competition between diffusion and the surface energy [57]. If there is sufficient anisotropy then the growth form leads to "symmetric" dendritic growth [Fig. 4(b)]. From a mathematical perspective, this instability is a consequence of the non-linear contributions to the equations of motion, which convert the spreading of the crystallization pattern from a diffusive ($R \sim t^{1/2}$) to wavelike ($R \sim t$) propagation. Physically, this dramatic increase in front speed results from a drastic reduction in distance that the solute rejected by the interface must diffuse, because the liquid channels between the fingers act as a local solute sink. The highly enriched liquid is thus incorporated (trapped) into the growing crystal, in accord with the Keith and Padden picture of spherulitic growth. Such structures are also obtained in eutectic crystallization, where the second phase plays an analogous role to the liquid channels [58–61].

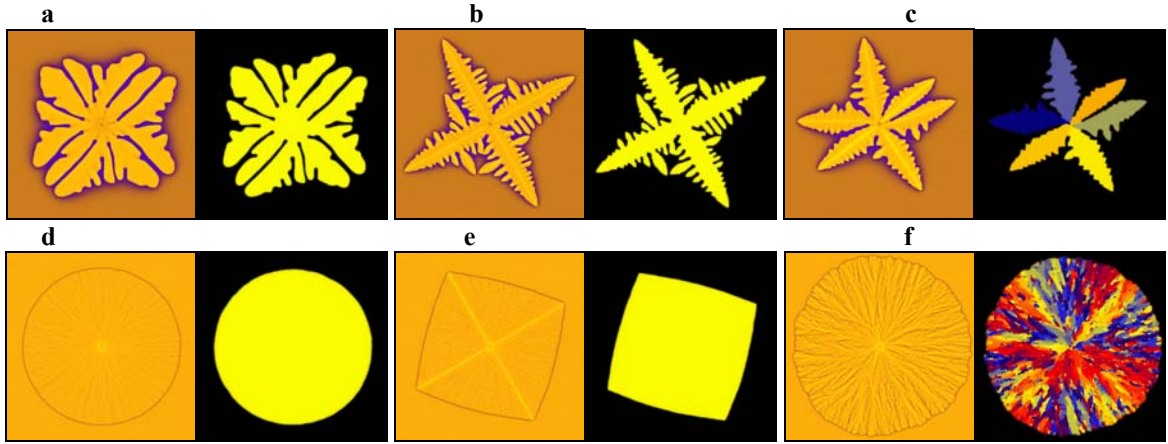

FIG. 4. From single crystals to category 1 spherulites. Single crystal growth forms for (a), (d) isotropic ($s_0 = 0$) and (b), (e) anisotropic interfacial free energy ($s_0 = 0.1$). (c), (f) Polycrystalline morphologies obtained by repeating the anisotropic calculations while reducing the orientational mobility by a factor of 0.15. Composition maps (odd columns) and orientation maps (even columns) are shown. All calculations were performed on a $500 \times 500$ grid (6.6 μm × 6.6 μm). Computations performed at two supersaturations are presented (upper row: $S = 0.8$; lower row: $S = 1.0$). The phase field mobility is assumed isotropic. No metastable branching orientation is offered [the orientational free energy has only a single minimum ($x = 0$)]. Crystallization was initiated by inserting a slightly supercritical fluctuation at the center of the simulation window without orientational preference. The final crystallographic orientation develops from the fluctuating local orientation as determined by the governing equation. Since the same random noise was used in all cases, the 'yellow' direction nucleated when single crystals formed. In contrast, several orientations nucleated simultaneously, when reducing the orientational mobility. (Coloring: Composition maps: yellow – solidus, dark blue – liquidus. Orientation maps: When the fast growth direction is upwards, 30, or 60 degrees left, the grains are colored blue, yellow or red, respectively, while the intermediate angles are denoted by a continuous transition among these colors. Owing to the four-fold symmetry, orientations that differ by 90 degree multiples are equivalent.)

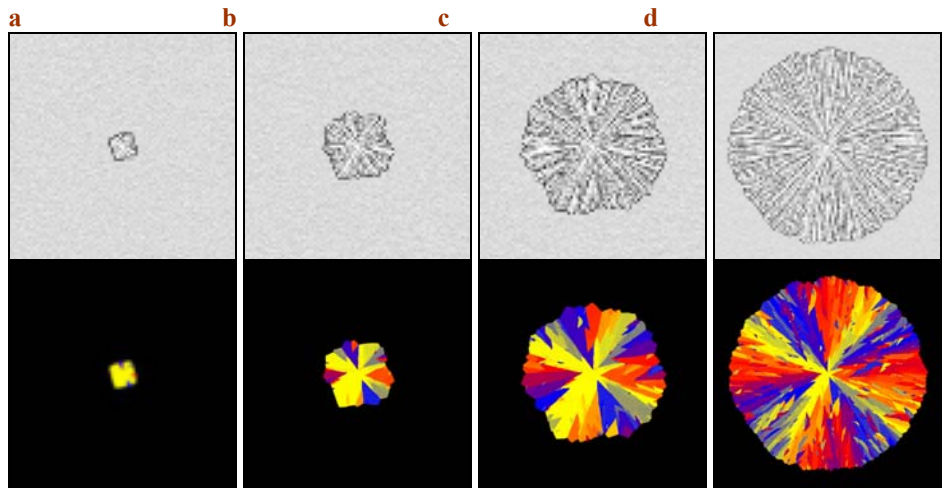

FIG. 5. Isothermal transition between a square-shaped single crystal and a category 1 spherulite induced by growth front nucleation, as predicted by the phase field theory. Note the gradual morphological transition, and the lack of a sharp demarcation line between areas solidified with square and spherulitic morphology in the fully-grown spherulite. With increasing size, the shape becomes more isotropic due to the randomizing effect of the newly formed grains. Note also the self-organized selection of grains whose maximum growth direction is perpendicular to the interface, yielding a cross-like pattern of grains with equivalent crystallographic orientations. [$4000 \times 4000$ grid. Snapshots taken at 1000, 2500, 5000, and 13 500 dimensionless time-steps, respectively are displayed. Panels (a)-(d) show the central $2,000 \times 2,000$ section of the simulation, while panel (d) shows the full $4000 \times 4000$ simulation.] Upper row: composition maps (a grayscale colormap was employed to increase contrast: black − liquidus, white − solidus). Lower row: orientation maps (coloring as in Fig. 4).

Spherical crystallization patterns also arise when "solute trapping" occurs, as manifested by the absence of solute rejection at the liquid-solid interface. This phenomenon occurs when the diffusion length approaches the interface width so that chemical diffusion and associated morphological instabilities are suppressed. Examples of this basic effect are illustrated in Figs. 4(d) and 4(e), which show single crystals growing under efficient solute trapping (extreme supersaturation) with and without anisotropy. Such regular single-crystal patterns are relatively rare.

Most spherical growth patterns observed in nature are polycrystals. The disorder of these structures emerges via growth front nucleation, which leads to a randomization of the local crystallographic orientation while retaining isotropy at large scales. Regardless of what growth form is dictated by crystallographic symmetry, these spherical growth forms occur robustly if the disorder is sufficiently large. In our view, this is the essence of spherulite formation.

The transition from crystalline to polycrystalline growth is illustrated in Fig. 4. The supersaturation $S$ is 0.8 or 1.0 in the upper and lower rows, respectively. The transition between the symmetric dendrite shown in Fig. 4(b) and the polycrystalline dendrite in Fig. 4(c) occurs as $\chi$ is reduced to model the influence of dynamic heterogeneities [31]. If we additionally increase the supersaturation, we obtain a highly branched polycrystalline crystallization pattern with an average circular shape, as shown in Fig. 4(f). This is a spherulite of category 1. Note the radially elongated grain structure, forming due to the self-organized selection of grains that have their fast growth direction perpendicular to the interface. Polycrystalline spherulites thus form when the driving force is large and the orientational mobility is small, a situation characteristic of highly undercooled complex liquids.

Regardless of the imposed crystallographic symmetries (two-, four- and six-fold were investigated), polycrystalline spherulites form with the same general structure. The fineness of the needle-like internal structures of the spherulites increases with increasing supersaturation.

*Category 1* spherulites have also been seen to form from transient single crystal nuclei [62]. Our model captures the gradual transition from square-shaped single crystals to circular shape under isothermal conditions. As seen in simulation, square-shaped single crystals nucleate after an initial incubation period. After exceeding a critical size (that depends on the ratio $\chi$ of the rotational and translational diffusion coefficients), the growing crystal cannot establish the same crystallographic orientation along its perimeter. Thus new grains form by growth front nucleation [31] as described in the introduction. This process gradually establishes a circular perimeter for large particles (Fig. 5).

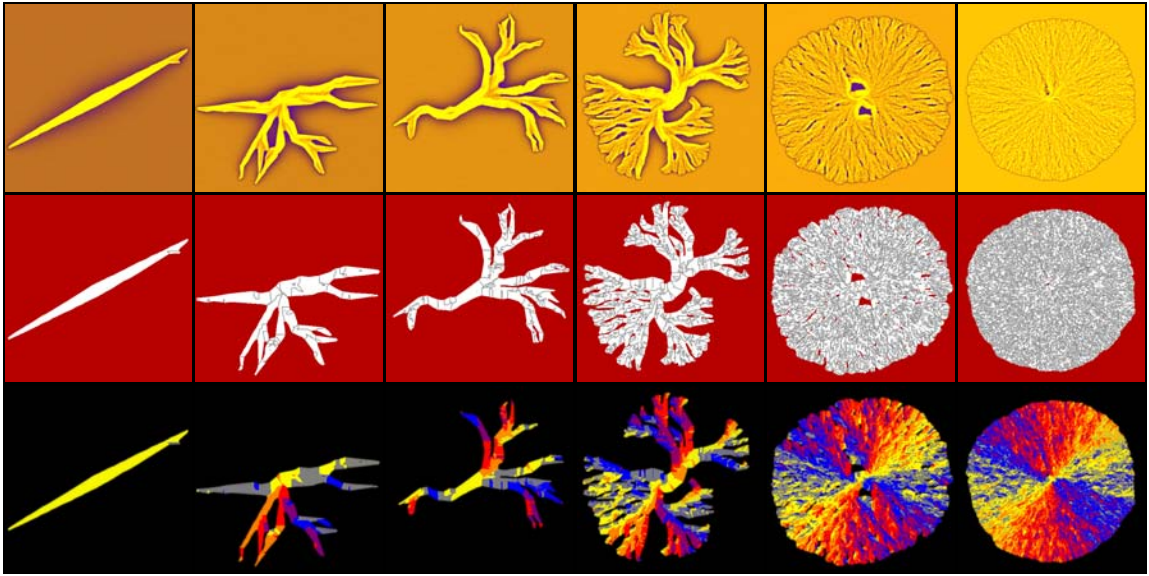

FIG. 6. Polycrystalline morphologies formed by random branching with a crystallographic misfit of 30 degrees. The kinetic coefficient has a two-fold symmetry and a large, 99.5%, anisotropy, expected for polymeric substances. Simulations were performed on a $500 \times 500$ grid (6.6 μm × 6.6 μm). Upper row: composition map (yellow − solidus, dark blue − liquidus). Central row: grain boundary map [gray scale in solid (crystal) shows the local orientational free energy density $f_{ori}$]. Lower row: orientation map. (The coloring of the orientation map is an adaptation of the scheme shown in previous figures for two-fold symmetry: When the fast growth direction is upwards, 60, or 120 degrees left, the grains are colored red, blue or yellow, respectively, while the intermediate angles are denoted by a continuous transition among these colors. Owing to two-fold symmetry, orientations that differ by 180 degree multiples are equivalent.) Unless noise intervenes, six different orientations are allowed, including the orientation of the initial single crystal nucleus, which was set common for all simulations [30 degrees off horizontal direction (yellow)]. In the present color code, yellow, gray, blue, purple, red, and orange stand for them. In order to make the arms better discernible, in the orientation map, the liquid (which has random orientation, pixel by pixel) has been colored black. The supersaturation varies from left to right as $S = 0.75$, 0.85, 0.90, 0.95, 1.00, and 1.10. Note the chain of transitions that links the needle-crystal forming at low supersaturation, to 'axialites', crystal 'sheaves', and eventually to the spherulites (with and without 'eyes' on the two sides of the nucleus).

Many studies of the early stages of spherulite growth, especially in polymers, indicate that these structures initially grow as slender thread-like fibers [15,21,30]. These structures successively branch to form space-filling patterns. We thus adopt a strong two-fold symmetry for the kinetic coefficient, ensuring fibrillar growth, and include a preferred misorientation angle of 30 degrees ($m = 3$ and $x = 0.15$). The resulting growth morphologies are shown as a function of supersaturation in Fig. 6. As in Fig. 4, the crystal evolves from a symmetric single crystal to a spherulite as the supersaturation is increased. We observe that with increasing driving force there is an increased branching frequency, yielding more space-filling patterns. Thus we obtain an array of patterns: fibrils, sheaves, spherulites with partially formed eyes, and fully developed *category 2* spherulites. We see (second row Fig. 6) that the 'eyes' become increasingly small with increasing supersaturation, due to the increase in GFN. The consequence of our imposed misorientation is evident in the third row of Fig. 6, where there are six preferred orientations, corresponding to the imposed 30 degrees misorientation preference. This effect is especially pronounced at low supersaturations, while at high supersaturations noise-driven faults randomize the local orientation.

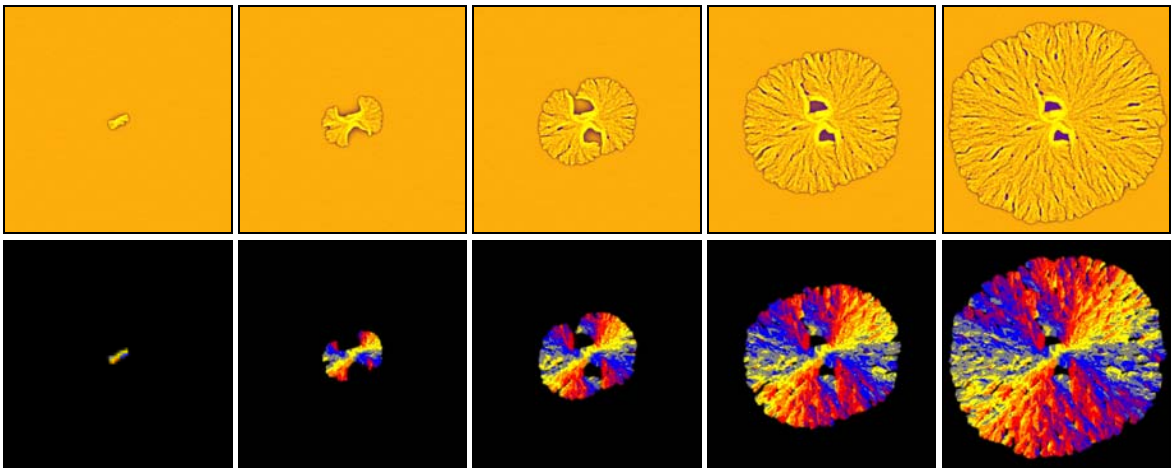

FIG. 7. The birth of a category 2 spherulite at $S = 1.0$, in the phase field theory. Time increases from left to right. (Snapshots taken at 4.2, 8.4, 12.6, 21, 33.5 μs after nucleation are shown. The dimensionless time used in the calculations has been transformed to real time using the diffusion coefficient of liquid Ni-Cu: $D_{NiCu} = 10^{-5}$ cm²/s. For other diffusion coefficients $D$, the times presented here have to be multiplied by $D_{NiCu}/D$.) Upper row: composition map; lower row: orientation map. Coloring and other conditions are as for fifth column in Fig. 6.

Next, the time evolution of a *category 2* spherulite is considered at a fixed supersaturation (Fig. 7). First, fibrils form and then secondary fibrils nucleate at the growth front to form crystal 'sheaves'. The diverging ends of these sheaves subsequently fan out with time to form eyes [Figs. 1(g) and 1(h)], and finally a roughly spherical growth form emerges. This progression of spherulitic growth is nearly universal in polymeric materials [15,21].

What characterizes the difference between category 1 and 2 spherulites? For category 1 spherulites, isotropy is achieved rapidly. In Fig. 4(f), we observe the initial crystal had a 4-fold symmetry, and the high frequency of GFN and the associated branching leads to isotropic growth. Thus, disorder disrupts the crystalline anisotropy early in the growth process, yielding category 1 spherulites. In Fig. 7 the initial growth is fibrillar, in contrast with Fig. 4, and it takes much longer, at the same level of supersaturation (and consequent GFN), for this randomization to occur. The occurrence of category 2 spherulites is directly related to the prevalence of early-stage fiber-type growth in comparison with the branched growth. In addition, as we increase the driving force, the time at which the growth becomes isotropic on average decreases and the structural differences between category 1 and 2 spherulites diminish.

We wish to specify in which systems these growth patterns are prevalent. Category 1 spherulites, are a normal mode of growth in metallic and mineral systems, where fibrous growth is relatively rare. On the other hand, category 2 spherulites are ubiquitous in polymeric systems. In such fluids, high supercoolings are readily attained due to their complex molecular structure, and the fiber growth habit is characteristic of the chain-folding mechanism by which polymers crystallize [12–21,63–65].

Category 1 and 2 spherulites may form under the same experimental conditions. How can this be understood? The early stage of growth strongly influences the late stage morphology of the spherulite. Under circumstances where the initial growth form is perturbed by fluctuations, an admixture of category 1 and 2 spherulites is obtained. For example, simultaneous nucleation of several orientations within the same nuclei should generally yield category 1 spherulites, but such events may be rare, and so the structures will coexist with category 2 spherulites. Such multiorientation nucleation events have been found in experiments on silica embedded silver particles [66] and by atomistic simulations for simple liquids [67]. Multiple nucleation events have been observed in atomic force microscopy measurements of polymer spherulite formations in thin films [68–70].

## B. Transformation kinetics

In the growth of compact space-filling spherulites chemical or thermal diffusion plays a negligible role. Under these conditions, the time evolution of the extent of crystallization $X$ follows the Johnson-Mehl-Avrami-Kolmogorov (JMAK) scaling

$$X = 1 - exp\{-[(t - t_0)/\tau]^p\}, \qquad (3.1)$$

where $t_0$ is an incubation time due to the relaxation of the athermal fluctuation spectrum, $\tau$ is a time constant related

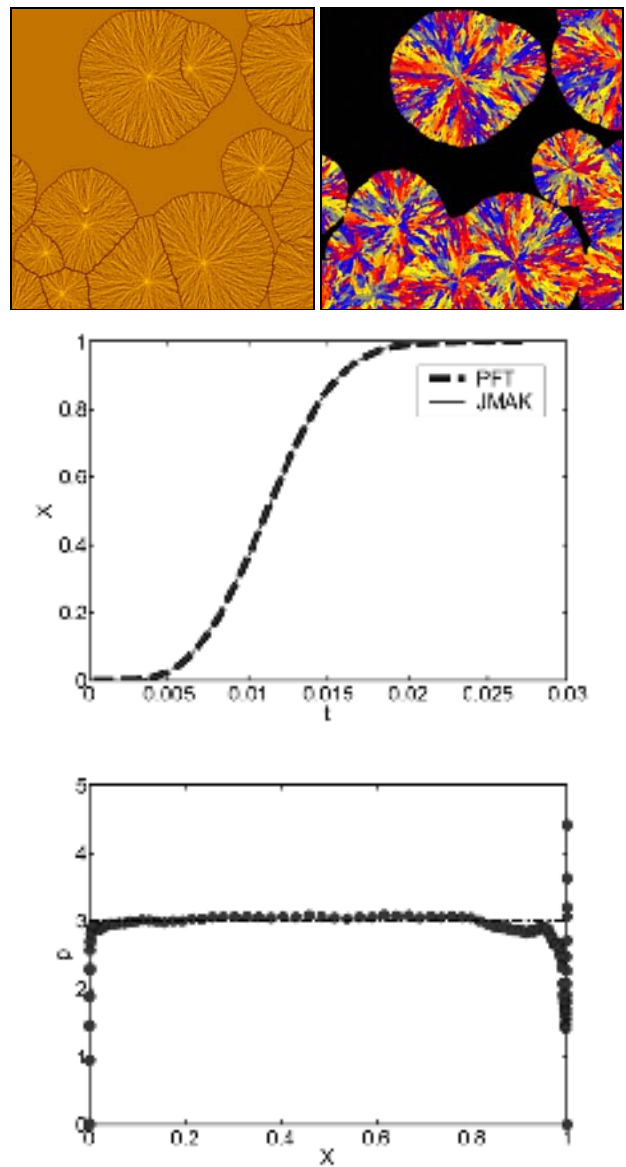

FIG. 8. Nucleation and growth of polycrystalline spherulites in the phase field theory. ($1000 \times 1000$ section of a simulation on a $5000 \times 5000$ grid). Upper row: Left concentration map; right orientation map. Central row: Transformed fraction vs. dimensionless time (dashed line), JMAK curve with the best-fit parameters (solid line). Bottom: Kolmogorov exponent as a function of crystalline fraction.

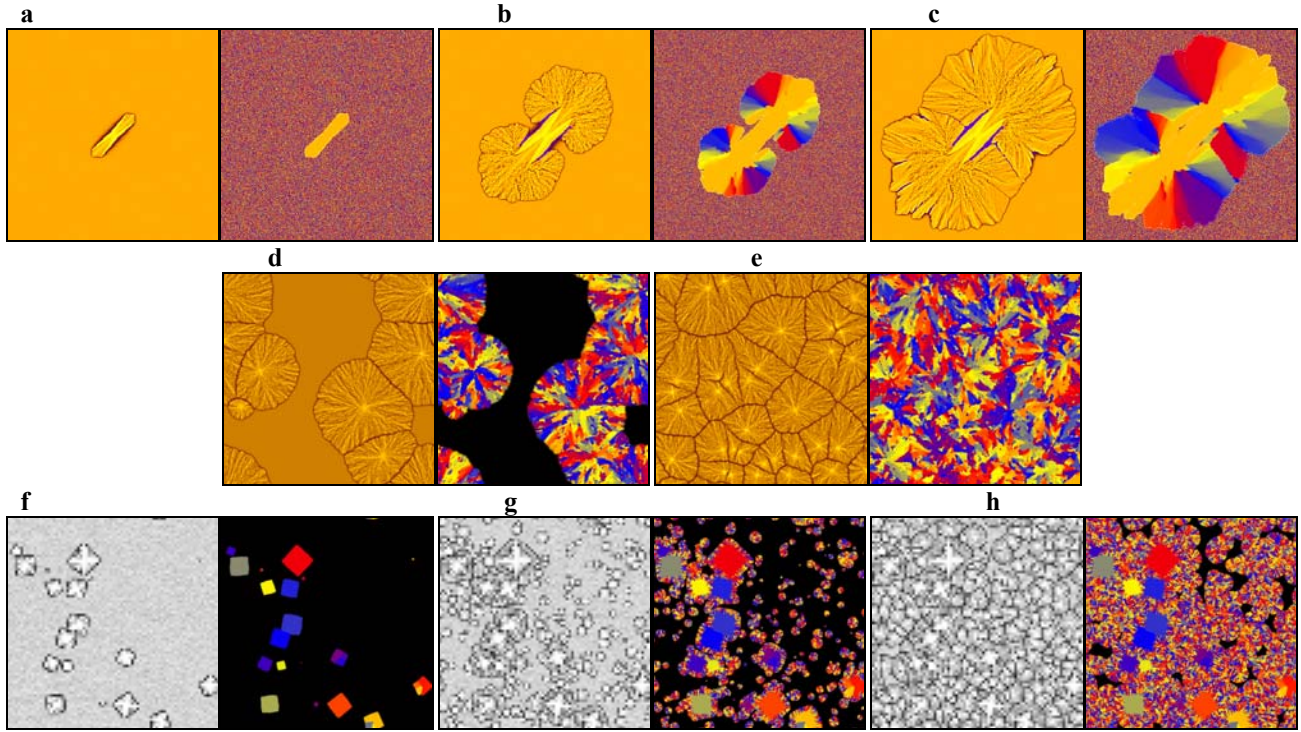

FIG. 9. Multistage heat treatments involving spherulitic solidification, as predicted by the phase field theory: (a) – (c) Transition between a faceted crystal habit [(a), nucleated at 1575 K] and (b) a spherulitic array after the sample is quenched to, and crystallized isothermally at 1571 K ($M_0$ is reduced by a factor of 20), and back to faceted growth (c) after returning to 1575 K. (Compared to polymers, this system requires a relatively small temperature cycling range due to the ideal solution behavior of the Ni-Cu system.) Note the formation of new crystal grains due to GFN during the low temperature stage of the cycling. The computations were performed with 5% anisotropy of the interface free energy (of six-fold symmetry) and 85% anisotropy of the phase field mobility (of two-fold symmetry) on a $1000 \times 1000$ grid. In (d) and (e) we show two thermal histories with the same final temperature. In (d) spherulitic solidification occurs at 1574 K after direct quenching from above the melting point (1595 K). In (e) spherulitic solidification occurs at 1575 K, after deep quenching first to 1350 K. Note the similarity of the growth forms, and the enhanced number of crystallites in the latter case. The computations were performed with 10% anisotropy of the interface free energy (of four-fold symmetry) on a $500 \times 500$ grid. In (f)–(h) spherulitic overgrowth occurs on pre-existing square crystals with parallel nucleation and growth of spherulites. Square-crystals were formed at 1574 K isothermally, then quenched to 1564 K where crystallization completed. The computations were performed with 15% anisotropy of the interface free energy (of four-fold symmetry) on a $1,000 \times 1,000$ grid. Left: composition maps (coloring: blue – liquidus, yellow – solidus, except the last row, where a grayscale colormap was employed to increase contrast: black – liquidus, white – solidus). Right: orientation maps (coloring as in Figs. 6 and 7 of the paper).

to the nucleation and growth rates, and $p = 1 + d$ is the Kolmogorov exponent, while $d$ is the number of dimensions [71]. This relationship is exact if (i) the system is infinite; (ii) the nucleation rate is spatially homogeneous; and (iii) either a common time-dependent growth rate applies or anisotropically growing convex particles are aligned in parallel (for derivation of Eq. (3.1) by the time cone method, see Refs. 72 and 73). For constant nucleation and growth rates in an infinite 2D system $p = 3$ applies. We investigated the transformation kinetics for noise-induced nucleation under the conditions shown in Fig. 4(f) for a relatively large system ($5000 \times 5000$ grid). To avoid the unnatural starting transient emerging from noiseless initial conditions (constant phase and concentration fields), first we heat-treated the system at 1595 K (above the liquidus curve) for 10 000 time steps, then we quenched it to 1574 K. The results are shown in Fig. 8. Fitting Eq. (3.1) to the simulation data between $0.01 < X < 0.95$ (where the data are the least noisy), we find $p = 3.04 \pm 0.02$ (and $\tau = 0.0106$

$\pm 0.00005$, $t_0 = 0.00178 \pm 0.00005$), which is reasonably close to the $p = 3$, expected for such a transition [71].

## C. Multi-step heat treatments

There is a great deal of interest in how temporal variations in processing conditions (temperature, pressure, etc.) influence spherulitic growth morphology. Multistage heat treatments on polymeric substances have demonstrated that that both the local growth morphology and growth rate depend on the temperature, but are independent of previous thermal history [1,10,62]. For example, cycling between two temperatures reversibly switches between faceted and spherulitic growth morphologies both in experiment [1,10] and simulation [Figs. 9(a)-(c)]. The predominance of either growth morphology depends on the cycling time, and complex patterns are generated in this fashion. For example, following experiment [62], we can simulate either a direct quench to the temperature of spherulitic solidification from above the melting point or instead simulate a deeper quench followed by heating to the same final temperature. As shown in Figs. 9(d),(e) these different histories yield much the same late stage growth form, but a larger number of spherulites in the latter deep quench case (due to enhanced nucleation at lower temperatures). Finally, other experiments [62] show that spherulitic overgrowth occurs on square-shaped crystals grown at small undercoolings, while, simultaneously, normal spherulites fill the remaining space. This behavior is recovered by our phase field simulations [Figs. 9(f)–(h)]. The ability of this theory to reproduce such complex sequences suggests that our field theory contains the essential physics necessary to describe a broad range of real materials.

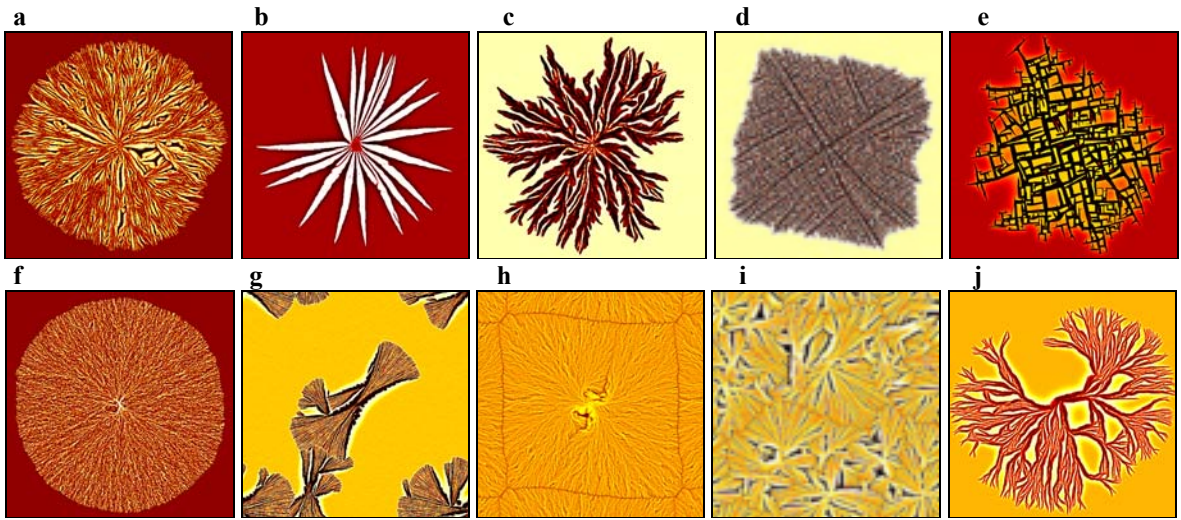

FIG. 10. Sherulitic morphologies as predicted by the phase field theory. The contrast of the composition maps was changed to enhance the visibility of the fine structure. Compare the predicted morphologies to the patterns in Fig. 1. The kinetic or interfacial free energy anisotropies have a two-fold symmetry in all cases; other conditions for these simulations are presented in Table II.

### D. Morphological variability

We now return to the wide range of spherulitic crystallization patterns shown in Fig. 1. Can the current model explain this variability? Fig. 10 shows a selection of simulations that bear resemblance to the morphologies displayed in Fig. 1. In addition to the category 1 and 2 spherulites described above, we observe structures ranging from spiky and arboresque spherulites, to 'quadrites' [15,16] exhibiting a cross-hatching fine structure [see Fig 1(d)], to undulating branched patterns. These simulations differ only in the driving force, anisotropies, branching angle, and mobilities, indicating that the essential features of a broad variety of spherulitic morphologies can be captured, using only a few model parameters (Table II). We note that while the anisotropy for the interfacial free energy ($s_0$) in some of these calculations significantly exceeds the missing orientation threshold (1/3, for twofold symmetry), we do not expect the results to be qualitatively different if this issue is addressed through a convexification approach such as that of Eggleston et al. [74].

### E. Spherulites vs duality of static and dynamic heterogeneities

In a recent paper [31], we have shown that particulate additives and quenched-in orientational disorder may lead to similar growth morphologies and morphological transitions. Examples of the analogous roles played by static and dynamic heterogeneities (foreign particles and quenched-in orientational defects) in spherulitic growth are displayed in Fig. 11: The conversion of a single crystal spherulite into polycrystalline ones, and the transition between a needle-shape single crystal and a densely branched spherulitic morphology, termed loosely as "fungus" are shown. The foreign particles are represented by orientation pinning centers, an economical description developed in Ref. 45. These examples show that the duality outlined in Ref. 31 is valid also for the spherulitic structures. Whether this remains so during the multi-stage heat treatments is uncertain, and needs further investigation.

### IV. DISCUSSION

Spherulite formation arises from a variety of mechanisms that lead to nucleation at the crystallization growth front. Heterogeneities, either static or those intrinsic to supercooled liquids, result in growth front nucleation of new grains and associated branching of the growing crystal. We have modeled the effects of these dynamic heterogeneities by appropriately reducing $\chi$, the ratio of the rotational to translational mobilities. Ultimately, this growth front nucleation (otherwise termed 'symphathetic' [24], 'double' [25–28], or 'secondary nucleation' [63–65]) randomizes the local crystallographic orientation, leading at long times to structures having an isotropic (spherical and circular in 3D and 2D, respectively) average large-scale structure. The variety of spherulites derives from the variability in the crystallographic symmetries of the parent crystal, the rate at which thermal fluctuations cause the crystallization front to branch, constraints on the orientation of newly formed grains and ordinary side-branching initiating from the growing dendritic tips. This competition between these basic processes leads to a rich variety of spherulitic patterns that are captured by our phase field modeling.

TABLE II. Conditions for simulations shown in Fig. 10.

| Panel | $S$ | $s_0$ | $\delta_0$ | $x$ | $\varphi$ | $m_{\phi,0}$ | $m_{\theta,0}$ | $N$ |
|-------|-----|-------|------------|-----|-----------|--------------|----------------|-----|
| (a) | 1.0 | 0.335 | 0.75 | 0.1 | 30° | 1.0 | 144 | 1000 |
| (b) | 0.75 | 0.5 | 0.5 | 0.0 | - | 0.9 | 108 | 2000 |
| (c) | 0.95 | 0 | 0.995 | 0.275 | 15° | 1.0 | 360 | 1000 |
| (d) | 0.9 | 0 | 0.995 | 0.15 | 90° | 1.0 | 360 | 2000 |
| (e) | 0.9 | 0.75 | 0.995 | 0.15 | 90° | 1.0 | 1440 | 500 |
| (f) | 1.1 | 0 | 0.995 | 0.2 | 30° | 1.0 | 360 | 1000 |
| (g) | 0.9 | 0.5 | 0 | 0.0 | - | 0.9 | 360 | 2000 |
| (h) | 1.1 | 0.5 | 0 | 0.2 | 15° | 1.0 | 360 | 500 |
| (i) | 0.9 | 0.5 | 0 | 0.0 | - | 0.8 | 180 | 1000 |
| (j) | 0.875 | 0 | 0.995 | 0.2 | 15° | 1.0 | 18 | 3000 |

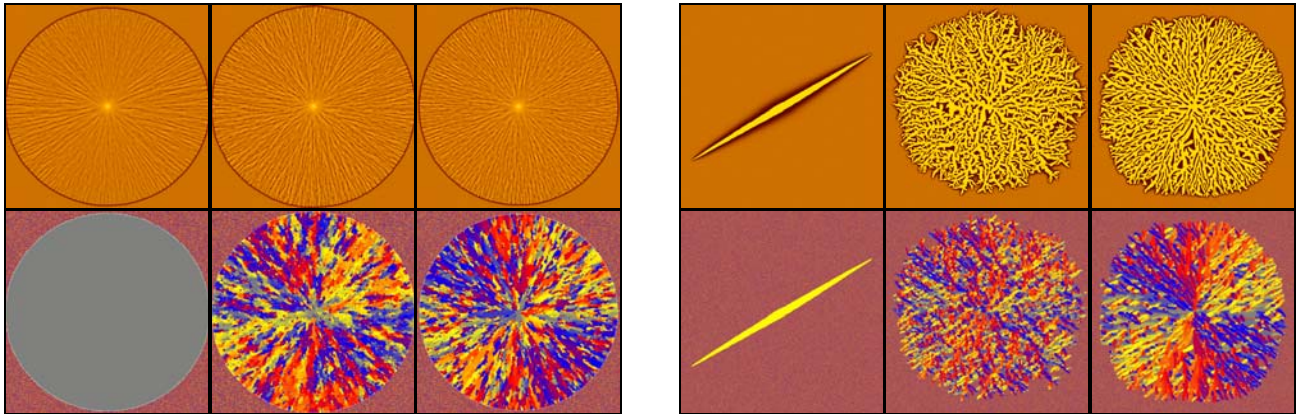

FIG. 11. Spherulitic structures and the duality of static and dynamic heterogeneities in the phase field theory: *Block on the left:* Single crystal spherulite (left column) and polycrystalline spherulites produced by introducing foreign particles (center) and by reducing the orientational mobility (right), respectively. The orientational mobility is the same for all but the third column, where it has been reduced by a factor of 4. There are 15,000 five pixels sized foreign orientation pinning centers that have been introduced into the simulations shown in the second column. The calculations were performed on a $1,000 \times 1,000$ grid (13.2 μm × 13.2 μm). The interface free energy is assumed isotropic while the anisotropy of the phase field mobility is 5 %. *Block on the right:* Single crystal needle (left column) and polycrystalline spherulitic "fungi" produced by introducing foreign particles (center) or by reducing the orientational mobility (right). The orientational mobility is the same for all but the third column, where it has been reduced by a factor of 5. $N = 250,000$ single-pixel-sized orientation pinning centers have been introduced into the simulation shown in the second column. The calculations were performed on a $2,000 \times 2,000$ grid (26.4 μm × 26.4 μm). The interface free energy is assumed isotropic while the anisotropy of the phase field mobility is 99.5%, and has a two-fold symmetry ($k = 2$). Coloring as for Fig. 8.

It is remarkable that our coarse-grained model yields such morphological diversity, given its reliance such a small set of thermodynamic and transport properties. We have simply assumed the existence of a first order phase transition coupled to chemical diffusion, and included thermal fluctuations and a description of the underlying crystalline anisotropies. The molecular-scale details are captured by bulk physical properties, such as the surface free energy anisotropy, diffusivities, etc. Despite the minimal nature of the model, we are able to reproduce much of diversity and structural complexity of spherulitic growth forms.

## V. SUMMARY

We have presented a phase field theory that incorporates diffusional instabilities, the trapping of orientational disorder due to reduced rotational diffusional coefficient, and random crystallographic branching. We have demonstrated that our model:

(1) Describes well the formation of category 1 and 2 spherulites;
(2) Yields the proper transformation kinetics;
(3) Reproduces morphological changes, seen in multi-stage heat treatments;
(4) Captures the morphological variability of the spherulites with only a limited number of model parameters.

Extension of the treatment to other complex polycrystalline morphologies is underway.

## ACKNOWLEDGMENTS

L. G. thanks Mathis Plapp and Tamás Börzsönyi for the many enlightening discussions on the phase field theory, and to Tamás Börzsönyi for his earlier contributions to the development of the model. This work has been supported by contracts OTKA-T-037323, ESA PECS Contract No. 98005, and by the EU Integrated Project IMPRESS, and forms part of the ESA MAP Projects No. AO-99-101 and AO-99-114. T. P. acknowledges support by the Bolyai János Scholarship of the Hungarian Academy of Sciences.